\documentclass{article}

\usepackage{PRIMEarxiv}

\usepackage[utf8]{inputenc} 
\usepackage[T1]{fontenc}    
\usepackage{hyperref}       
\usepackage{url}            
\usepackage{booktabs}       
\usepackage{amsfonts}       
\usepackage{nicefrac}       
\usepackage{microtype}      
\usepackage{lipsum}
\usepackage{fancyhdr}       
\usepackage{graphicx}       
\graphicspath{{media/}}     
\usepackage{authblk}
\usepackage{amssymb}
\usepackage{lipsum}
\usepackage{amsmath} 
\usepackage{adjustbox}
\usepackage{arydshln}
\usepackage{graphicx}
\usepackage{braket}
\usepackage{longtable}
\usepackage{soul}
\usepackage[T1]{fontenc}
\usepackage{color}
\usepackage[dvipsnames]{xcolor}
\usepackage{subcaption}
\usepackage{booktabs}
\usepackage{textcomp}
\usepackage{stackrel}
\usepackage[normalem]{ulem}

\usepackage[english]{babel}
\title{A Review on Practical Challenges of \\Aerial Quantum Communication
}

\author[1]{Umang Dubey}
\author[1]{Prathamesh Bhole}
\author[2,3]{Arindam Dutta}
\author[1]{\\Dibya Prakash Behera}
\author[1]{Vethonulu Losu}
\author[1]{Guru Satya Dattatreya Pandeeti}
\author[1]{\\Abhir Raj Metkar} 
\author[1]{Anindita Banerjee}
\author[3]{Anirban Pathak}
\affil[1]{Centre for Development of Advanced Computing, C-DAC Innovation Park, Pune- 411008, Maharashtra, India}
            
\affil[2]{Department of Physics, Indian Institute of Technology, Jodhpur- 342030, Rajasthan, India}

\affil[3]{Jaypee Institute of Information Technology, A-10, Sector-62, Noida- 201309, Uttar Pradesh, India} 

\begin{document}
\maketitle

\begin{abstract}
The increasing demand for the realization of global-scale quantum communication services necessitates critical investigation for a practical quantum secure communication network that relies on full-time all-location coverage. In this direction, the non-terrestrial quantum key distribution is expected to play an important role in providing agility, maneuverability, relay link, on-demand network, and last-mile coverage. In this work,  we have summarized the research and development that has happened until now in the domain of quantum communication using non-terrestrial platforms with a specific focus on the associated challenges and the relevant models. Further, to extend the analysis beyond the existing know-how, a hybrid model involving the features of Vasylyev \textit{et al.}'s model and Liorni \textit{et al.}'s model is introduced here. The hybrid model entails us adapting a spherical beam to an elliptic beam approximation and effectively capturing the characteristics of transmittance in densely humid weather conditions and at low altitudes. Further, to understand the potential impact of the weather conditions of a region on atmospheric attenuation, as an example the average monthly visibility of Pune city was analyzed for the years 2021 and 2022. In addition, a simulation of a generic model is performed using a software-defined network paradigm where quantum teleportation is simulated between distant parties using a swarm of drones in NetSquid.
\end{abstract}


\section{Introduction}
\label{introduction}
Quantum communication offers a fundamentally secure way to establish long-distance communication channels, making it highly relevant for secure communication in critical applications where traditional encryption methods may be vulnerable to future quantum attacks. Quantum communication has many facets and the two most important facets are secure quantum communication and teleportation, both are unique in some sense, teleportation does not have any classical analog, and quantum cryptography can be unconditionally secure whereas classical cryptography can never achieve that feature. 

\begin{figure}[h]
    \centering
    \includegraphics[width=\textwidth]{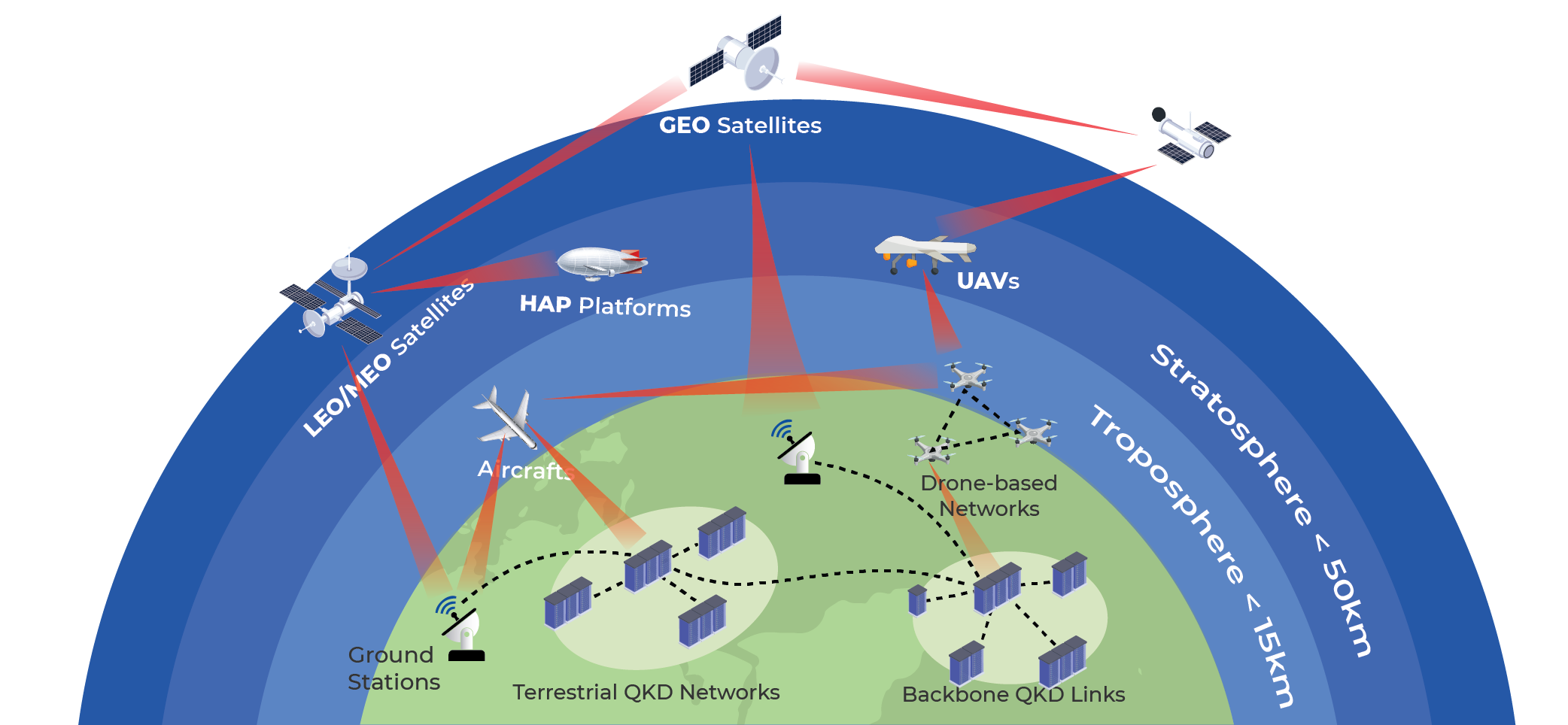}
    \caption{(Color online) Concept of aerial quantum communication \cite{aqkd-review}}
    \label{fig:concept}
\end{figure}

Quantum key distribution (QKD) is one of the cornerstones of quantum cryptography. It is a method of exchanging symmetric keys among parties, by leveraging the principles of quantum mechanics to ensure provable security against adversaries. Fiber and free-space are the most commonly used transmission mediums for QKD. However, there are several challenges in establishing practical and secure networks. These challenges include device imperfections, such as detector noise, polarization-extinction-ratio, and signal loss depending on the transmission medium. In fiber-based QKD, the losses increase significantly with the distance, making it unfeasible over larger geographical areas. Free-space QKD offers the advantage of extended coverage and flexibility but is susceptible to losses caused by atmospheric turbulence, fog, and other environmental factors in the communication channel \cite{P21, P+21}. Satellite-based QKD is considered a potential candidate for long-distance communication, however, along with the free-space propagation challenges, it faces a limited operational timing window, non-agility, and higher infrastructural costs. These factors collectively impede achieving higher key rates in satellite-based QKD systems. However, to realize a practical quantum secure communication network that would ideally provide full-time all-location coverage, all the modes of transmission need to function in an integrated fashion. Here, the utilization of aerial platforms \cite{aqkd-review} may offer a highly flexible, cost-effective, and re-configurable approach for expanding the reach of quantum communications across time and space. In Fig. \ref{fig:concept}, we have illustrated the concept of aerial quantum communication, with a hierarchical quantum network operating in different atmospheric layers. Deploying of aerial quantum nodes such as drones,  high altitude platforms (HAPs), hot-air balloons, unmanned aerial vehicles (UAVs), and aircraft can serve as temporary relays. It can also act as intermediate mobile nodes between terrestrial ground stations and satellites and can be used for resolving the last-mile quantum key exchange challenge for inner-city or field networks due to their rapid deployment capabilities. Moreover, for higher altitudes, low-velocity aircraft can provide longer link duration and broader transmission coverage.

\begin{table*}[h!]
\centering
\begin{tabular} {l l l l l l l l }
\hline \textbf{Year} & \textbf{Distance} & \textbf{Secure} & \textbf{$\lambda$} & \textbf{Pulse repe-} & \textbf{QKD} & \textbf{QBER} & \textbf{Demonstration} \\ 
& (km) & \textbf{key rate} & (nm) & \textbf{-tition rate} & \textbf{protocol} & & \\ \hline
    1989 & 30cm & - & - & 403 bits & - & 66 bits & On table at IBM \cite{10.1145/74074.74087} \\ \hline
    1992 & 32cm & - & - & 217 bits & - & 2-4\% & Free air optical path \cite{Bennett1992} \\ \hline
    1997 & 0.205 & 50 Hz & 772 & - & B92 & 1-6\% & Over indoor paths \cite{PhysRevA.57.2379} \\ \hline
    1998 & $\sim 1$ & 3.5-45 KHz & 772 & 10 MHz & B92 & 1.5 \% (D) & Los Alamos (D) \\ 
    & & & & & & 2.1\% (N) & National Laboratory (N) \cite{1998APS..4CS..C104B} \\ \hline
    2002 & 9.81 & 50.78 Kb (D) & 772 & - & BB84 & 5\% (D) & Los Alamos Ski Club,\\ 
         &      & 118.06 Kb (N) &  &   &      & 2.1\% (N) & The National Forest Service \cite{Hughes_2002} \\ \hline
    2002 & 23.4 & 1.5-2 Kbps & - & - & BB84 & 5\% & Tx-Zugspitze, South Germany \cite{Kurtsiefer2002}\\ 
         & & & & & & & Rx- Mountain of Karwendelspitze \\ \hline
    2004 & 0.73 & 1 Mbps & 845 & 250 ps & B92 & 1.1\% & Free-space \cite{Bienfang:04}\\ \hline
    2004 & 13 & 10 bps & 702 & - & BB84 & 5.83\% & Tx - Dashu Mountain\\ 
    & & & & & & & Hefei of China (elevation- 281 m) \\ 
    & & & & & & & Alice-West Campus of USTC\\ 
    & & & & & & & Bob-Feixi of Hefei \cite{PhysRevLett.94.150501} \\\hline
    2006 & 144 & 417 bits & 710 & 249 MHz & BB84 & 4.8\% & La Palma and Tenerife \cite{Ursin2007} \\ \hline
    2006 & 1.5 & 850 bps & 404 & - & BB84 for & 5.4\% & Free-space \cite{10.1063/1.2348775} \\ 
    & & & & & pol. ent. p. & & \\ \hline
    2006 & 0.48 & 50 Kbps & 850 & - & BB84 & 3-5\% & Free space, Munich \cite{munich-2006} \\ \hline
    2007 & 144 & 12.8, 42 bps & 850 & 10 MHz & DS BB84 & 6.48\% & La Palma and Tenerife \cite{PhysRevLett.98.010504} \\ \hline
    2008 & 1.575 & 85 bps & 815 & - & BBM92 & 4.92\% & Free-space \cite{Erven:08} \\ \hline
    2008 & $\sim 1.5$ & 300 bps & 407 & - & Modified E91 & $\sim$ 3\% & Free-space \cite{Ling_2008} \\
    & & & -810 & & & & \\ \hline
    2010 & 1.305 & 2.7 Kbps & 404 & - & BBM92 & 2.48 & Free-space \cite{10.1007/978-3-642-11731-2_14} \\ \hline
    2013 & 20 & 7.9 bps & 850 & 10 MHz & BB84 & 4.8\% & Dornier 228 turboprop aircraft\\ 
    & & & & & & & and the optical ground station \cite{Nauerth2013} \\ \hline
    2013 & $\sim 96$ & 159.4 bps (MP) & 850 & 100 MHz & DS & 4.04\% & MP: Over a turntable \\
         &           & 48 bps (FP) & & & & & FP: Hot-air balloon \cite{Wang_2013} \\ \hline
    2014 & 600 & 100 Kb & - & 76 MHz & DS & 4.3-5.51\% & QEYSSAT- 600 km\\ 
         &     & & & & & & altitude microsatellite \cite{10.1117/12.2041693} \\
         \hline
    2014 & 2.5-7.5 & - & 850 & 1 MHz & BB84 & - & Tx: Helicopter (100 kmph) \\ 
     & & & & & & & Rx: Top floor of a building\\ 
     & & & & & & & in an airport \cite{Zhang-2014} \\ \hline
    2015 & $\sim$ 0.650 & 40 bps & 532, & 80 MHz & DS BB84 & 6.16\% & Pickup truck \\
    & & & 810, & & & & traveling at 33 kmph \\
    & & & 1550 & & & & angular speed \cite{Bourgoin:15} \\ \hline
    2017 & 1200 & 1.1 Kbps & 850 & 100 MHz & DS BB84 & 1-3\% & Micius- 635 kg satellite \cite{Liao_2017} \\ \hline
    2017 & 802 & $\sim$ 10-100 bps & 800 & 10 MHz & B92 & < 5\% & SOCRATES- 50 kg \\ 
    & & & & & & & microsatellite \cite{Takenaka2017} \\ \hline
    2017 & 3-10 & 868 Kb & 785 & 400 MHz & DS BB84 & 3-5\% & Twin Otter- research aircraft \cite{Pugh_2017}\\ \hline
    2017 & - & - & 650 &  500 KHz & DS BB84 & - & On table (towards DJI S1000+ \\ 
    & & & & & & & octocopter QKD) \cite{Drone-2017} \\ \hline
    2021 & 0-0.04 & - & - & - & BB84 & $\sim$ 50\% & Amov- lab's Z410 drone \\ 
    & & & & & & & with T- engine 2216 \\ 
    & & & & & & & and Pixhawk flight control QKD \\ \hline
    2022 & 30 cm & 4 - 15.3 kbps & 850 & 100 MHz & BB84 & 2.4\% & Hand-held sender \cite{PhysRevApplied.18.024067} \\ \hline
    2023 & 0.2 & 8 KHz & 850 & 50 MHz & BB84 & 2.22-2.32\% & Drone-QKD \cite{drone-bb84-2023} \\ \hline
    2021- & $10^a$ & - & 850 & 50 MHz & 3 states & 2.22-2.32\% & a. Drone-Drone: DJI S1000+\\
    2023 & & & & & & & drone to Alta 8 Pro drone \cite{Drone-2017, Drone-2021}  \\
    & & & & & & & b. Drone-Car \cite{Drone-2021, 10.1117/12.2647923}\\
    & & & & & & & c. Car-Car \cite{Drone-2021, 10.1117/12.2647923} \\ \hline 
\end{tabular}
\caption{Developments towards aerial quantum communication around the world, where $\lambda$: Wavelength, QBER: Quantum bit error rate, D: Day, N: Night, DS: Decoy state, pol. ent. p.: polarization-entangled photons, MP: Moving platform, FP: Floating platform, Tx: Transmitter, Rx: Receiver} 
\label{Table-I}
\end{table*} 

Present-day drones, or UAVs, encompass a wide spectrum of capabilities, spanning take-off weights ranging from a few grams to several tons. They can operate at cruising altitudes that vary from a few meters above the ground to altitudes exceeding 20 kilometers. Furthermore, their flight duration can extend up to 25 days. Considering these recent advancements it is imperative to consider these UAVs to establish mobile quantum networks (QNs), enabling on-demand and real-time coverage across diverse spatial and temporal scales. This will enable quantum communication \cite{china-ed} from distances of kilometers (local-area networks) to hundreds of kilometers (wide-area networks). This approach represents a flexible and economically viable means of expanding the reach of secure communication while delivering real-time coverage as needed. 

Several works have been reported in this area, which includes air-to-ground QKD demonstration using the Dornier-228 aircraft by Nauerth \textit{et al.} \cite{Nauerth2013}, downlink QKD demonstration using the hot-air balloon by Wang \textit{et al.} \cite{Wang_2013}, the basis detection and compensation experiment using the Z-9 helicopter by Zhang \textit{et al.} \cite{Zhang-2014}, the free-space QKD based on a moving pick-up truck by Bourgoin \textit{et al.} \cite{Bourgoin:15}, uplink QKD demonstration using the Twin Otter research aircraft by Pugh \textit{et al.} \cite{Pugh_2017}, the drone-based QKD test using DJI S1000+ octocopter by Hill \textit{et al.} \cite{us-2021} and drone-based entanglement distribution using UAV by Liu \textit{et al.} \cite{china-ed, china-relayed-2021}. The work by Liu \textit{et al.} laid the foundations for establishing re-configurable mobile QNs. Recently drone-based QKD, with an average secure key rate larger than 8 kHz using decoy-state BB84 protocol with polarization encoding was demonstrated \cite{drone-bb84-2023}. There have been a few demonstrations of the satellite QKD also, including a B92 protocol implementation \cite{Takenaka2017} using the SOCRATES (Space Optical Communications Research Advanced Technology Satellite), and a 600 km DS-QKD implementation \cite{10.1117/12.2041693} using the QEYSSAT microsatellite. In Table \ref{Table-I}, we have reported the developments in aerial quantum communication to date. 

Considering the fact that aerial QKD is emerging as a potential candidate for the efficient implementation of a practical secure quantum communication network. It is interesting to address the implementation challenges and their impact on the performance of aerial QKD systems. Consequently, in Section \ref{tech-chall}, the technological challenges are presented in detail. In Section \ref{hybrid-model}, we introduce a hybrid model for low-altitude communication that takes into account real-world scenarios. In Section \ref{section-4}, we discuss the link configurations, budgeting, and margin in detail, along with time synchronization. Section \ref{aqns} presents the simulation of quantum teleportation using a swarm of drones based on quantum software-defined networking (QSDN) oriented architecture. Finally, the paper is concluded in Section \ref{conc}. 
\section{Technological challenges}
\label{tech-chall}
There are several challenges associated with the implementation of aerial quantum communication. One of the major challenges in achieving long-distance aerial quantum communication is the loss of signal in the transmission medium, this can be caused due to various physical reasons. Before we describe them, we may note that in an optical fiber, the losses increase exponentially with the length of the fiber and it is denoted by the attenuation coefficient ($\beta_{\rm a}$), expressed in dB/km. It depends on the fiber material, manufacturing tolerances, and wavelength. It is about 2 dB/km at 800 nm, 0.35 dB/km at 1310 nm, and 0.2 dB/km at 1550 nm.  Secure quantum communication is usually done through telecom-grade optical fiber using light of wavelength about 1550 nm, where the attenuation is minimum at $\sim$0.2 dB/km. It can be slightly reduced further by using ultra-low-loss-fiber with a nominal attenuation coefficient of 0.158 dB/km and that can increase the distance for quantum key distribution to some extent. However, to perform secure quantum communication, beyond a few hundred km, one would be required to use a free-space route. Now, we may note that below fiber-based optical communication using light of wavelength below 800 nm is unusable as attenuation due to Rayleigh scattering increases considerably. Here appears an interesting point: there exists a high transmission window for free-space communication at around 770 nm. It is weakly dispersive and essentially non-birefringent at these wavelengths. This provides a great advantage to free-space communication. However, free-space transmission has some drawbacks, too. Particularly, its performance depends on the atmospheric conditions. For example, the transmission of the signal through a turbulent medium may lead to arrival time-jitter, beam wander, beam pointing error, beam divergence, etc. In this section, we will systematically discuss the technological challenges that arise due to these issues with a specific focus on how to model the effect of atmospheric conditions. To begin with we may discuss the effect of atmospheric turbulence.

\subsection{Atmospheric turbulence}
Air turbulence \cite{laserna} in the atmosphere plays a significant role in free-space optical (FSO) communication as it can affect the operating laser beam, leading to beam divergence, beam wandering, scintillation, etc. Several efforts have been made to mathematically describe the effect of atmospheric turbulence on the FSO \cite{FSOAir}. One such effort led to the development of energy cascade theory \cite{kolmogorov_1962}.

The energy cascade theory is a fundamental concept in the study of turbulence in the Earth's atmosphere. It explains  energy transferred from large-scale turbulent motion to smaller and smaller scales. It states that the outer scale eddy $L_{o}$, and inner scale eddy $l_{o}$, form the bounds of an inertial sub-range. The eddies in the inertial range are statistically homogeneous and isotropic. Within this range, large eddies break into smaller eddies transferring energy. This process carries on until inner scale eddy $l_{o}$ is reached. After this, energy dissipates through viscosity. In 1940s, Andrey Kolmogorov \cite{Kolmogorov_1941} obtained a beautiful expression for the wavenumber spectrum (now known as the Kolmogorov spectrum) in the turbulence inertial subrange. The Kolmogorov spectrum describes the refractive index fluctuations as

\begin{equation}
    \phi_{n}(k)=0.033 C_{n}^2k^\frac{-11}{3},      \frac{1}{L_{o}}<<k<<\frac{1}{l_{o}}
    \label{eq:kol-spec-ref}
\end{equation}
where, \textit{k} is the wavenumber, and $C_{n}^2$ is the refractive index structure parameter.

The refractive index variations arise due to changes in temperature and pressure with varying altitudes. The refractive index structure constant, $C_{n}^2$, is a parameter used to characterize refractive index of air variations thus, the strength of air turbulence. It has values ranging from $10^{-17}m^\frac{-2}{3}$ to $10^{-13}m^\frac{-2}{3}$ to describe weak to strong turbulence, respectively \cite{yan}. It serves as a valuable tool for assessing both the scintillation index and the Rytov variance.

\begin{figure}[h!]
\centering
\begin{subfigure}{0.3\textwidth}
\includegraphics[width=\textwidth]{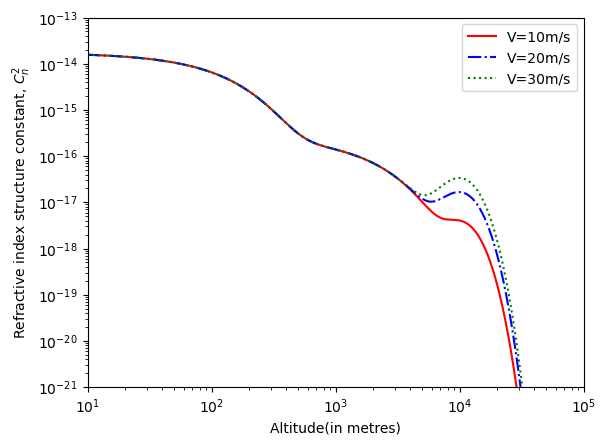} 
\caption{}
\label{fig:HVB}
\end{subfigure}
\hfill
\begin{subfigure}{0.3\textwidth}
\includegraphics[width=\textwidth]{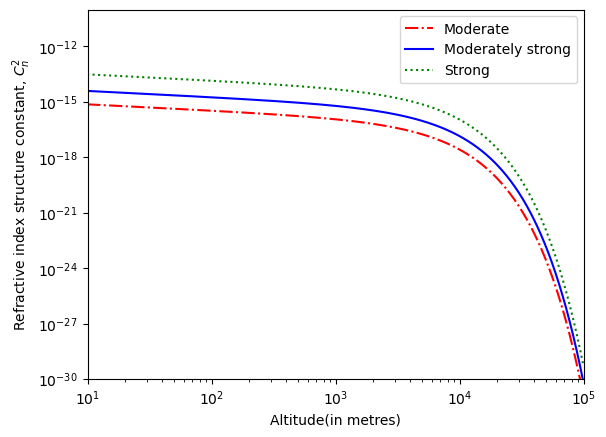}
\caption{}
\label{fig:fried-model}
\end{subfigure}
\hfill
\begin{subfigure}{0.3\textwidth}
\includegraphics[width=\textwidth]{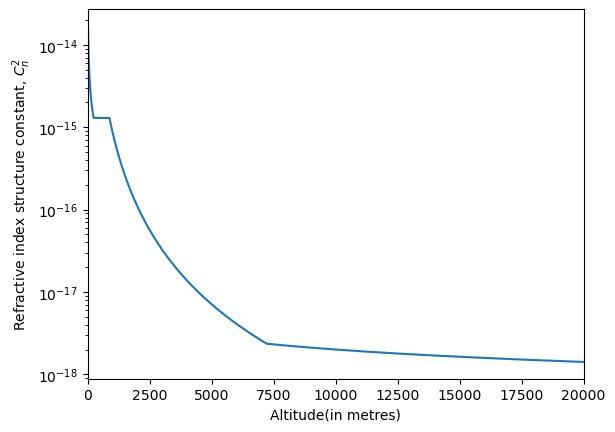}
\caption{}
\label{fig:SLC-D}
\end{subfigure}
\caption{Plot for structure parameter constant $C_{n}^2$ with altitude (a) using HVB model with varying velocities (b) using Fried model for  moderate, moderately strong, and strong conditions and (c) using SLC model.}
\label{fig:image2}
\end{figure}

Certain models offer a means to depict the impact of atmospheric turbulence on $C_{n}^{2}$ \cite{LC06}. Among these, the Hufnagel-Valley Boundary (HVB) model \cite{zilber} is used for long-range propagation. The model incorporates various on-site conditions such as wind speed, iso-planatic angle, and altitude. Using the HVB model, $C_{n}^2$ was plotted for different wind velocities as shown in Fig. \ref{fig:HVB}. Higher wind velocities have higher $C_{n}^2$ values depicting a highly turbulent atmosphere. Fried \cite{Fried-65} proposed another model for determining $C_{n}^2$. It is valid for only short-range propagation. For the Fried model, $C_{n}^2$ was plotted using the turbulence strength parameter, $K_{o}$ values for strong, moderately strong, and moderate conditions are shown in Fig. \ref{fig:fried-model}. $C_{n}^2$ increases for increasing $K_{o}$ values showing more turbulent environments. Further, an alternative model used for describing the refractive index structure constant at low altitudes is known as Submarine Laser Communications-Day (SLC-D) \cite{GBM+88}. This model is particularly well-suited for use in environments characterized by high humidity, dense cloud cover, and underwater conditions (refer to Appendix A for detailed information). It is shown in Fig. \ref{fig:SLC-D} gives a daytime turbulence profile and is also well suited for long-range propagation.

\subsubsection{Scintillation and beam wandering}
Atmospheric turbulence affects the propagation of the optical beams leading to wavefront distortions. It can cause fluctuations in the intensity of the beam, such that we obtain speckled patterns on the beam wavefront at the receiver end. This phenomenon is known as scintillation. It occurs because the turbulent atmosphere causes different parts of the beam to experience varying refractive index gradients. Scintillation causes loss in signal-to-noise ratio and deep signal fades. Aperture averaging \cite{Churnside:91} is one of the techniques used to mitigate scintillation. 

\begin{figure}[h!]
    \centering
    \includegraphics[width=0.5\textwidth]{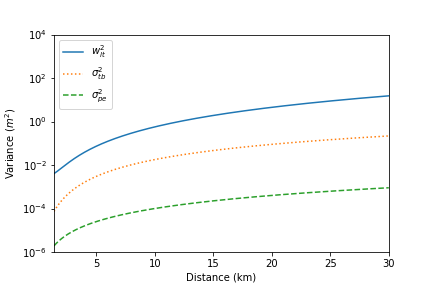}
    \caption{(Color online) Variance $\sigma_{pe}^2, \sigma_{tb}^2$ and $w_{lt}^2 $ for varying distances.}
    \label{fig:dist_vs_var}
\end{figure}

Beam wandering arises as a result of two distinct factors: atmospheric turbulence along the path of the beam and random errors in the transmitter's pointing mechanism. These two factors operate independently and their effects accumulate over the course of propagation. When transmitting an optical signal through free space, one observes the random displacement of the instantaneous centroid of the signal, often referred to as the "hot spot" or point of maximum irradiance. This quivering, which is assumed to follow a Gaussian distribution with variance $\sigma^2$, is commonly known as beam or centroid wandering. In essence, this wandering phenomenon is a consequence of both pointing error, denoted as $\sigma^2_{pe}$, stemming from Gaussian jitter and off-target tracking, and atmospheric turbulence, represented by $\sigma^2_{tb}$. These two effects are mutually independent, and their combined effect results in the total variance of wandering, denoted as $\sigma^2 = \sigma^2_{pe} + \sigma^2_{tb}$ \cite{Ghalaii2022}. The impact of $\sigma^2_{pe} \text{ and } \sigma^2_{tb}$ varies depending on the different weather conditions, wavelength used, beam size and shapes, etc. In Fig. \ref{fig:dist_vs_var}, variance of the beam centroid wandering resulting from turbulence ($\sigma^2_{tb}$), pointing error ($\sigma^2_{pe}$) and the long-term beam waist ($w^2_{lt}$) are plotted for $\lambda =$ 800 nm and initial radius of collimated beam  $w_{0}=$ 5 cm a. It is observed that $w_{lt}^2\gg \sigma_{tb}^2\gg\sigma_{pe}^2$ for all distances. The parameters $w_{lt}^2, \sigma_{tb}^2$ and $\sigma_{pe}^2$ are shown to have a logarithmic growth with increasing distance. Other parameters are the outer scale of turbulence $L_{0} = 1$ m and $C^2_n = 1.28$x$10^{-14} m^{-2/3} $ (night-time operation).\\ 

\subsubsection{Atmospheric attenuation}
Signal loss and link failure are caused by atmospheric attenuation due to absorption, scattering, and scintillation. All these effects vary with time and depend on the current local conditions, weather, and distance. The atmospheric attenuation $(\tau)$ in dB for distance $L$ (km) and $\beta_{\rm a}$ attenuation coefficient, can be given by:
\begin{equation}
    \tau =4.3429 \beta_{\rm a} L
    \label{atm-atten}
\end{equation}
The absorption loss is mainly due to the carbon dioxide molecules and water particles, whereas the scattering loss is due to the snow, fog, clouds and rain present in the atmosphere. For weather conditions such as clear weather to dense fog weather, scattering loss varies between 0.21 dB/km to 0.84 dB/km \cite{kim}. It can be characterized as follows:

\textbf{Attenuation coefficient due to fog and rain:} Attenuation due to scattering of the optical signal depends on the visibility range of the link. And the visibility varies depending on different weather conditions. The attenuation factor for different weather conditions such as fog and rain is given by:
\begin{equation}
   \beta_{\rm fog}= \left( \frac{3.91}{V} \right) \left(\frac{\lambda}{550}\right)^{-p}
   \label{att-fac-fog}
\end{equation}

\begin{equation}
   \beta_{\rm rain}=\left(\frac{2.8}{V}\right)
   \label{att-fac-rain}
\end{equation}
where, \emph{V} (km) is the visibility and \emph{p} is the size distribution coefficient of scattering.

\begin{figure}[h!]
\centering
\begin{subfigure}{0.4\textwidth}
\includegraphics[width=\textwidth]{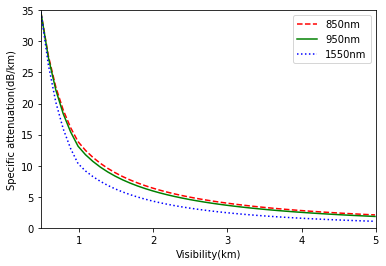} 
\caption{ }
\label{fig:thickfog}
\end{subfigure}
\hfill
\begin{subfigure}{0.4\textwidth}
\includegraphics[width=\textwidth]{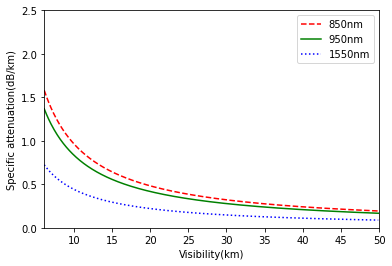}
\caption{ }
\label{fig:lightfog}
\end{subfigure}
\caption{Specific attenuation vs visibility using wavelengths 850 nm, 950 nm and 1550 nm which are frequently used in FSO communication for (a) thick fog condition and (b) light fog and haze condition.}
\label{fig:image3}
\end{figure}

Attenuation for thick fog, light fog, and haze conditions can be modeled by the Kim \cite{Kim-model} or Kruse \cite{Kruse} model. Kim model is able to describe attenuation for visibility less than 1 km. For thick fog conditions where visibility is under 0.5 km, $\emph{p}=0$,  Thus, attenuation is the same for all operating wavelengths. As visibility increases, the attenuation reduces overall. Higher wavelength values have slightly less attenuation when compared to lower wavelength values. See Fig. \ref{fig:thickfog} and Fig. \ref{fig:lightfog} to visualize the effect of fog. 

Size distribution, \emph{p} are chosen depending on the visibility range as defined in the Kruse and Kim models. According to the Kim model, 
\begin{equation}
    p =  
    \begin{cases}
      1.6  & \text{ when } V > \text{50 km}\\
      1.3 & \text{ when } \text{6 km} < V < \text{50 km}\\
      \text{0.16 }V \text{+ 0.34} & \text{ when } V < \text{6 km} \\
      V \text{ - 0.5} & \text{ when } \text{0.5 km} < V < \text{1 km} \\
      0 & \text{ when } V < \text{0.5 km.} \\
    \end{cases}
\end{equation}
According to the Kruse model,
\begin{equation}
    p =  
    \begin{cases}
      1.6 & \text{ when } V > \text{50 km}\\
      1.3 & \text{ when } \text{6 km} < V < \text{50 km}\\
      \text{0.585 }V^{1/3} & \text{ when } V < \text{6 km.} \\
    \end{cases}
\end{equation}

\begin{figure}[h!]
\centering
    \includegraphics[width=0.5\textwidth]{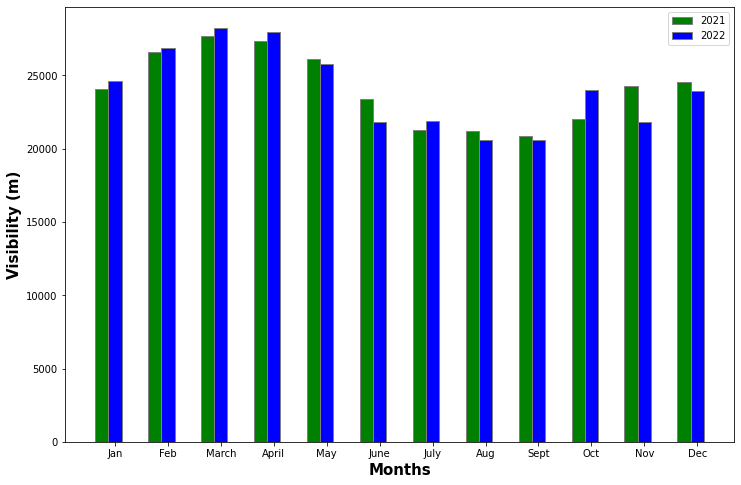}
    \caption{(Color online) Comparison of average monthly visibility for Pune city for the years 2021 and 2022.}
    \label{fig:visibility-vs-Months}
\end{figure}

We have investigated the average visibility of Pune city for the last two years using the data collected from the Indian Meteorological Department (IMD) (refer to Fig. \ref{fig:visibility-vs-Months}). Pune city is chosen just as an example, as we plan to perform experimental aerial quantum communication in Pune. We observe that due to the changing weather conditions of any region, there are variations in the average visibility of the atmosphere. Therefore, the performance of any aerial quantum communication system would depend on the date and time when it's used. Additionally, the integration of weather monitoring systems and predictive algorithms can aid in optimizing system performance by adjusting parameters in response to changing weather conditions. Overall, understanding and mitigating the effects of weather and visibility is crucial for reliable aerial quantum communication.

\textbf{Atmospheric extinction}: An additional significant source
of signal loss during the free-space transmission of an optical beam
is atmospheric extinction. This phenomenon results from the combined
impact of aerosol absorption and Mie/Rayleigh scattering. When we
consider free-space communication at a constant altitude $\overline{h}$,
this phenomenon can be quantified using the straightforward Beer-Lambert
equation, $\eta_{{\rm atm}}\left(\overline{h}\right)=e^{-\alpha\left(\overline{h}\right)z}$,
where, $\alpha\left(\overline{h}\right)$ is the extinction factor
which varies depending on both the altitude and the wavelength of
the signal \cite{BH08}. Neglecting refraction, the atmospheric transmissivity
can be expressed as 
\begin{equation}
\eta_{{\rm atm}}\left(\overline{h},\phi\right)={\rm exp}\left\{ -\int_{0}^{z\left(\overline{h},\phi\right)}dx\,\alpha\left[\overline{h}\left(x,\phi\right)\right]\right\} ,\label{eq:AD4}
\end{equation}
while taking into consideration a generic zenith angle ($\phi$). 

\textbf{Atmospheric transmittance:} Atmospheric transmittance is a measure of the amount of incoming electromagnetic radiation (such as visible light, infrared, or microwave radiation) that passes through the Earth's atmosphere without being absorbed, scattered, or otherwise attenuated. Different wavelengths of electromagnetic radiation are affected differently as they pass through Earth's atmosphere. The variation in transmittance with wavelength is primarily due to the absorption and scattering properties of the atmospheric constituents, like gas molecules, aerosols, etc., at different wavelengths.

\begin{figure}[h!]
\centering
    \includegraphics[width=0.6\textwidth, height=6cm]{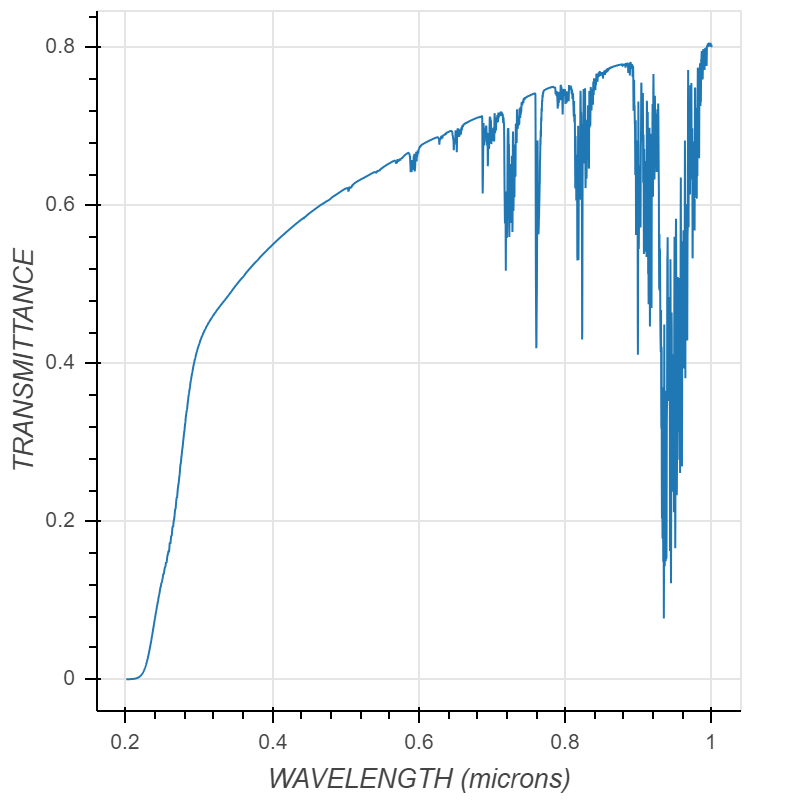}
    \caption{Simulated atmospheric transmittance for zenith with different wavelengths.}
    \label{fig:tx-vs-lambda}
\end{figure}

In Fig. \ref{fig:tx-vs-lambda}, we have presented a simulation for the atmospheric transmittance for a 1 km FSO link as a function of different wavelengths along the zenith for the downlink configuration was carried out using the MODTRAN software, which was developed by the Spectral Sciences Inc. (SSI) and the Air Force Research Laboratory of The United States of America (USA) for an urban location with the tropical atmospheric model and 9 km visibility.

The results provide an indication for the identification of the optimum wavelengths necessary for the free-space link establishment like the APT coarse and fine-tracking laser beams, entangled pair distribution, and time synchronization.


\subsubsection{Beam divergence loss}
\label{Diff-loss}
One of the major sources of loss in establishing a point-to-point link, with accuracy for single mode fiber (SMF) coupling (where the SMFs typically have the mode field diameter of around 5 $\mu m$), is the diffraction-induced beam broadening.

The optical beam propagation through the atmosphere spreads out owing to the diffraction, leaving the receiver with a narrow field of view (FOV), not being able to collect a fraction of the transmitted power, resulting in the beam divergence loss, also known as the geometric loss. 

One may consider the Gaussian beam as a quasi-monochromatic optical mode source with wavelength $\lambda$, and employ it for achieving free-space quantum communication. If this beam travels a distance of $z$, due to diffraction the spot size of the beam, $w_{D}$ will become: 
\begin{equation}
w_{D}=w_{0}\sqrt{\left(1-\frac{z}{R_{0}}\right)^{2}+\left(\frac{z}{z_{R}}\right)^{2}}\label{eq:AD3}
\end{equation}
where, the initial beam spot size is $w_{0}$ (smaller than the aperture
of the transmitter), radius of curvature is $R_{0}$, and ($z_R = \frac{\pi w_o^2}{\lambda} $)
is the Rayleigh length\footnote{For collimated Gaussian beam $\left(R_{0}=\infty\right)$, and consequently,
the spot size can be considered as $w_{D}=w_{0}\sqrt{1+\left(\frac{z}{z_{R}}\right)^{2}}$.}. Only a fraction of the initial beam is detectable and this fraction is determined by the diffraction-induced transmissivity, 
\begin{equation}
    \eta_{D}(z)=1-e^{-\frac{2a_{r}^{2}}{w_{D}^{2}}}
    \label{diff-trans}
\end{equation}
which may be approximated as, 
\begin{equation}
    \eta_{D}\simeq\eta_{D}^{{\rm far}}:=\frac{2a_{r}^{2}}{w_{D}^{2}}\ll1
    \label{approx-diff-trans}
\end{equation}
where $a_{r}$ is the aperture of the receiving telescope and $w_{D}$ spot size of the beam. 

Employing the PLOB (Pirandola-Laurenza-Ottaviani-Banchi) bound \cite{PLOB17} with the transmittance, we can estimate the upper bound of the maximum number of secret bits that can be distributed by a QKD protocol across a free-space communication channel by
\begin{equation}
    \mathcal{U}\left(z\right)=\frac{2}{\ln2}\frac{a_{r}^{2}}{w_{D}^{2}}
    \label{upper-bound}
\end{equation}
bits per use. 

\begin{figure}[h!]
\begin{minipage}[b]{0.45\linewidth}
\centering
    \includegraphics[width=\textwidth]{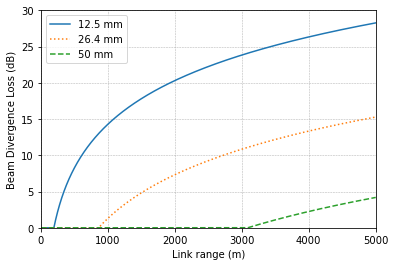}
    \caption{Beam divergence loss and the link distance for different transmitting and receiving optics aperture diameters.}
    \label{fig:diffloss-dist}
\end{minipage}
\hspace{0.5cm}
\begin{minipage}[b]{0.45\linewidth}
\centering
    \includegraphics[width=\textwidth]{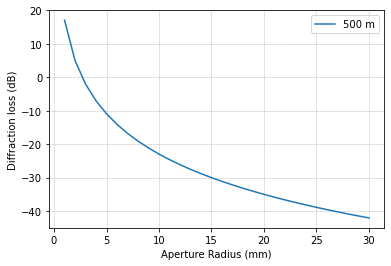}
    \caption{Beam divergence loss as a function of the beam aperture at a 500-m link distance (to be updated).}
    \label{fig:Diff-loss-vs-ba}
\end{minipage}
\end{figure}

Hence, it is important to choose the optimum transmitter and receiver optics aperture areas for the optimal beam diameters and low-pointing errors. Therefore, using Eq. (\ref{eq:beam-div-loss}), a simulation for the beam divergence loss, $L$ (dB) as a function of the diffraction-limited link distances within a local area network for small transmitting and receiving optics aperture diameters was carried out (refer to Fig. \ref{fig:diffloss-dist}),
\begin{equation}
    L \text{(dB)} = -10 \left[ \left(2\log{\frac{4}{\pi}}\right)+\log \left ({\frac{A_t A_r}{\lambda^2 z^2}}\right) \right]
    \label{eq:beam-div-loss}
\end{equation}
where,
$A_t$: aperture area of the transmitter optics, and $A_r$: aperture area of the receiver optics.

Similarly, the beam divergence loss as a function of the transmitter and receiver optics diameter at 500 m link distance is obtained (see Fig. \ref{fig:Diff-loss-vs-ba}). These results can aid in the identification of the proper transmitter and receiver optics aperture areas for the APT units to achieve longer link coverage, low pointing errors, and low diffraction-induced beam divergence loss.

It can be observed that the transmitter and receiver optics diameter of up to some centimeters, which can give the Rayleigh lengths of up to some hundreds of meters with low beam divergence loss are sufficient for the free-space communication within a local area mobile network. Further, an increase in the transmitting optics aperture area will effectively reduce the transmitter beamwidth, delivering the signal with more intensity, and hence reducing the beam divergence loss. However, it may lead to tight acquisition, pointing, and tracking requirements and will also increase the overall mass and the cost of the payload. 

Similarly, increasing the receiving aperture area scales the receiving signal power and reduces the beam divergence loss. However, it will also increase the collection of the amount of background noise by the receiver.  Therefore it implies that the effective performance improvement does not always scale linearly with the increasing transmitter and receiver optics aperture areas and an optimum choice needs to be made for the trade-off \cite{fsoc-book}. Also, for a long-distance link, we can reduce the effects of beam divergence loss by exploiting several shorter link segments and using the optical relay method \cite{china-relayed-2021} which is feasible, especially for drone-based platforms.

\begin{figure}[h!]
\centering
    \includegraphics[width=0.5\textwidth, height=5.5cm]{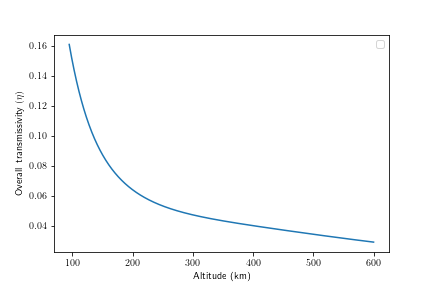}
    \caption{Altitude vs overall transmissivity}
    \label{fig:alt_vs_opt_loss}
\end{figure}

The overall transmissivity includes the multiplication of three types of optical transmissivity \cite{Ghalaii2022},

\begin{equation}
\eta = \eta_{D}\eta_{eff}\eta_{atm}    
\end{equation}
where, $\eta_{D}$ is turbulence or diffraction-induced transmissivity, $\eta_{eff}$ is receiver's efficiency and $\eta_{atm}$ is atmospheric loss. Overall transmissivity reduces with increasing altitude as shown in Fig. {\ref{fig:alt_vs_opt_loss}}.

Up to this point, we have delved into the significant and inevitable challenges faced in free-space quantum communication. These challenges encompass various factors, including atmospheric turbulence, scintillation, beam wandering, atmospheric attenuation, and beam divergence loss, all of which we have extensively discussed. In addressing these real-world effects, Vasylyev et al. introduced a model utilizing an elliptic beam approximation, demonstrating its strong compatibility with actual experimental data in their influential paper \cite{VSV+17, VSV16}. Liorni et al. extended this model for broader application in low earth orbit (LEO) satellite-based quantum communication \cite{LKB19} They factored in considerations such as the refractive index structure constant and the density of scattering particles, maintaining consistency with LEO satellite conditions, and evaluated their model under various weather scenarios. Now, our focus shifts to assessing the combined and realistic impact of these factors at lower altitudes, where communication can be facilitated using drones. To do this, we adapt their approach by incorporating the refractive index structure constant applicable to lower altitudes \cite{LC06, GBM+88} and introduce our hybrid methodology tailored for shorter altitude ranges.

\section{A hybrid model for low altitude signal transmission}
\label{hybrid-model}
In this section, we present a hybrid model using the model that exploits the properties of the Gaussian elliptical beam proposed by Vasylyev \textit{et al.}, \cite{VSV+17, VSV16}. Furthermore, we apply the generalized approach and incorporate day-time and night-time conditions, as introduced by Liorni \textit{et al.} in their seminal paper \cite{LKB19}. Their approach influences the transmittance value significantly, as transmittance relies on both beam parameters $\boldsymbol{V}$ and the diameter of the receiving aperture $a$. In order to enhance the readers' grasp of the elliptic beam approximation and its modified version, we provide a concise elucidation of the fundamental theory. A Gaussian beam is projected through a link that traverses both the atmosphere and a vacuum, originating from either a space transmitter (drone) or a ground station. This link is distinguished by its non-uniform characteristics. Typically, the changing intensity transmittance of this signal (the received beam) as it passes through a circular aperture with a radius $a_r$ in the receiving telescope is formulated as follows (see Refs. \cite{VSV16, VSV12} for details)

\begin{equation}
\eta = \int_{\left|\rho\right|^{2}=a_{r}^{2}}{\rm d^{2}\boldsymbol{\rho}\left|u\left(\mathbf{\boldsymbol{\rho}},\textit{z}\right)\right|^{2}}
\label{eq:Transmittance}
\end{equation}

In this context, the function $u\left(\mathbf{\boldsymbol{\rho}},z\right)$
represents the beam envelope at the receiver plane, which is situated
at a distance $z$ from the transmitter. The quantity $\left|u\left(\mathbf{\boldsymbol{\rho}},z\right)\right|^{2}$
signifies the normalized intensity concerning the entire $\boldsymbol{\rho}$
plane, where $\boldsymbol{\rho}$ denotes the position vector within
the transverse plane. The vector parameter $\boldsymbol{V}$ provides
a comprehensive description of the beam's state at the receiver plane
(see Fig. 1 in Ref. \cite{VSV+17}) and it's described as

\begin{equation}
\boldsymbol{V}=\left(x_{0},y_{0},W_{1},W_{2},\theta\right),\label{eq:Vector of beam-parameters}
\end{equation}
where $x_{0},y_{0}$, $W_{1}$, $W_{2}$ and $\theta$ represent the
coordinates of the beam centroid, the dimensions of the elliptical
beam profile (characterized by its principal semi-axes), and the orientation
angle of the elliptical beam, respectively. The transmittance is influenced
by these beam parameters in conjunction with the radius of the receiving
aperture ($a_{r}$).

In the context of an elliptical beam's interaction with a circular
aperture characterized by a radius denoted as $a_{r}$, the notion
of transmittance is precisely described by Equation (\ref{eq:Transmittance}).
The transmittance for this scenario can be articulated as follows \cite{VSV16}
\begin{multline}
\eta\left(x_{0},y_{0},W_{1},W_{2},\theta\right) = \frac{2\,\chi_{{\rm ex}t}}{\pi W_{1}W_{2}} \int_{0}^{a_{r}}\rho\,{\rm d}\rho\int_{0}^{2\pi}{\rm d}\varphi\,{\rm e^{-2A\left(\rho cos\varphi-\rho_{0}\right)}}{\rm e^{-2B\rho^{2}sin^{2}\varphi}}e^{-2{\rm C}\left(\rho{\rm cos}\varphi-\rho_{0}\right)\rho{\rm sin}\varphi}\label{eq:PDT Equation}
\end{multline}

Here, the symbol $a_{r}$ signifies the aperture's radius, while $\rho$ and $\varphi$ are used to express the polar coordinates of the vector
$\boldsymbol{\rho}$, we may write $x=\rho\,{\rm cos}\varphi$ and
$y=\rho{\rm \,sin}\varphi$, and $x_{0}=\rho_{0}{\rm \,cos}\varphi_{0}$
and $y_{0}=\rho_{0}\,{\rm sin}\varphi_{0}$, where $\rho_{0}$ and
$\varphi_{0}$ denote the polar coordinates associated with the vector
$\boldsymbol{\rho}_{0}$. Additionally, the expressions of the constants
are, ${\rm A}=\left(\frac{{\rm cos}^{2}\left(\theta-\varphi_{0}\right)}{W_{1}^{2}}+\frac{{\rm sin}^{2}\left(\theta-\varphi_{0}\right)}{W_{2}^{2}}\right),$
${\rm B}=\left(\frac{{\rm sin}^{2}\left(\theta-\varphi_{0}\right)}{W_{1}^{2}}+\frac{{\rm cos}^{2}\left(\theta-\varphi_{0}\right)}{W_{2}^{2}}\right),$
and ${\rm C}=\left(\frac{1}{W_{1}^{2}}-\frac{1}{W_{2}^{2}}\right){\rm sin\,2\left(\theta-\varphi_{0}\right).}$
Here, $\chi_{{\rm ext}}$ accounts for the influence of \emph{atmospheric
extinction}, which encompasses factors like back-scattering and absorption
that occur within the atmosphere \cite{BSH+13}.\textcolor{red}{{} }

With this elliptic beam approximation method one can relate the atmospheric
effect in free-space link at receiver's end. To make it more acceptable
and useful in real life situation, for free-space quantum communication,
Liorni \textit{et al.} proposed a generalized model \cite{LKB19}. Their model was generalized in the sense that it involved a non-uniform
link distribution between a drone and the ground, as described. To
calculate the moments of the distributions related to the parameters
of the elliptic Gaussian beam, we adopt the same Heaviside function
as employed in Liorni's model. We proceed to assess the expressions
for the first and second moments of the beam parameters ($\boldsymbol{V}$)
by making adaptations to Equations (4) through (9) from the Ref. \cite{LKB19}, aligning them with the conditions specific to drone-based
communication. We assume that the orientation angle $\theta$ of the
the elliptical profile follows a uniform distribution within the interval
$\left[0,\frac{\pi}{2}\right]$. In the context of up-links, the mean
value and variance of the beam's centroid position are consistent
in both the $x$ and $y$ directions and are equivalent to, $\left\langle x_{0}\right\rangle =\left\langle y_{0}\right\rangle =0,$
and $\left\langle x_{0}^{2}\right\rangle =\left\langle y_{0}^{2}\right\rangle =0.419\,\sigma_{R}^{2}w_{D}^{2}\Omega^{-\frac{7}{6}},$where
the\textcolor{red}{{} }term $\sigma_{R}=1.23\,C_{n}^{2}k^{\frac{7}{6}}z^{\frac{11}{6}}$\emph{
}is referred to as\emph{ Rytov parameter }which is an useful indicator
of integrated turbulence strength for extended propagation; $\Omega=\frac{k\,w_{D}^{2}}{2z}$
represents the Fresnel number, where $k$ denotes the optical wave
number and $w_{D}$ represents the beam spot size at the receiver.
In the chosen reference frame, the condition is set so that $\left\langle x_{0}\right\rangle =\left\langle y_{0}\right\rangle =0$.
The mean and (co)variance of $W_{i}^{2}$ can be expressed as, 

\[
\begin{array}{lcl}
\langle W_{i}^{2}\rangle & = & \frac{w_{D}^{2}}{\Omega^{2}}\left(1+\frac{\pi}{8}\,zn_{0}w_{D}^{2}+2.6\,\sigma_{R}^{2}\Omega^{\frac{5}{6}}\right),\vspace{2mm}\\
\langle\Delta W_{i}^{2}\Delta W_{j}^{2}\rangle & = & \left(2\delta_{ij}-0.8\right)\frac{w_{D}^{4}}{\Omega^{\frac{19}{6}}}\left(1+\frac{\pi}{8}\,zn_{0}w_{D}^{2}\right)\sigma_{R}^{2},
\end{array}
\]
where, $n_{0}$ denotes the scattering particles density\footnote{To estimate the value of $n_{0}$, which primarily comprises water
droplets, we utilize the atmospheric water vapor content profile.
This profile serves as our for understanding the scattering particles \cite{T84,TP88}}. Similar expressions are relevant for down-links when considering
the position of the elliptic beam centroid, $\left\langle x_{0}\right\rangle =\left\langle y_{0}\right\rangle =0,$
and $\left\langle x_{0}^{2}\right\rangle =\left\langle y_{0}^{2}\right\rangle =\alpha_{p}\,z,$also
the semi-axes of the elliptic beam profile are,
\[
\begin{array}{lcl}
\langle W_{i}^{2}\rangle & = & \frac{w_{D}^{2}}{\Omega^{2}}\left(1+\frac{\pi}{24}\,zn_{0}w_{D}^{2}+1.6\,\sigma_{R}^{2}\Omega^{\frac{5}{6}}\right),\vspace{2mm}\\
\langle\Delta W_{i}^{2}\Delta W_{j}^{2}\rangle & = & \left(2\delta_{ij}-0.8\right)\frac{3}{8}\,\frac{w_{D}^{4}}{\Omega^{\frac{19}{6}}}\left(1+\frac{\pi}{24}\,zn_{0}w_{D}^{2}\right)\sigma_{R}^{2},
\end{array}
\]
In this context, the symbol $\alpha_{p}\approx2$ $\mu{\rm rad}$ denotes
the approximate angular pointing error. Afterward, we employ the knowledge
of the probability distribution related to the elliptic beam parameters
(as expressed in equation \ref{eq:Vector of beam-parameters}) to
compute the probability distribution transmittance (PDT) using equation \ref{eq:PDT Equation} through a random sampling procedure using a
Monte Carlo methodology.

\begin{table}[h]
\centering
\begin{tabular}{c c c} 
\hline
 Parameter & Value & Description\\ 
\hline
$w_{D}$ & 1.15 cm & Down-link / up-link \vspace{1mm}\\
$a_{r}$ & 2.64 cm & Down-link / up-link \vspace{1mm}\\
$\lambda$ & 810 nm & Wavelength of the signal light \vspace{1mm}\\
$\beta$ & 0.7 & Parameter in $\chi_{{\rm ext}}(\phi)$ \vspace{1mm}\\
$\alpha_{p}$ & $2\times10^{-6}$ rad & Pointing error \vspace{1mm}\\
$\overline{h}$ & $18.5\,{\rm m}-240\,{\rm m}$ & Altitude of drone \vspace{1mm}\\
$n_{0}$ & 0.61 ${\rm m^{-3}}$ & Night-time condition \vspace{1mm}\\
$n_{0}$ & 0.01 ${\rm m^{-3}}$ & Day-time condition \vspace{1mm}\\
$C_{n}^{2}$ & $\frac{4.008\times10^{-13}}{\overline{h}^{1.054}}$  & Night-time condition \vspace{1mm}\\
$C_{n}^{2}$ & $\frac{3.13\times10^{-13}}{\overline{h}}$  & Day-time condition \vspace{1mm}\\
\hline \vspace{0.5pt}
\end{tabular}
\caption{Parameters linked to the optical and technical attributes of the transmission link with weather conditions.}
\label{Parameters-associated-with-link-length}
\end{table}


\begin{figure*}[h!]
\centering
\includegraphics[width=\textwidth]{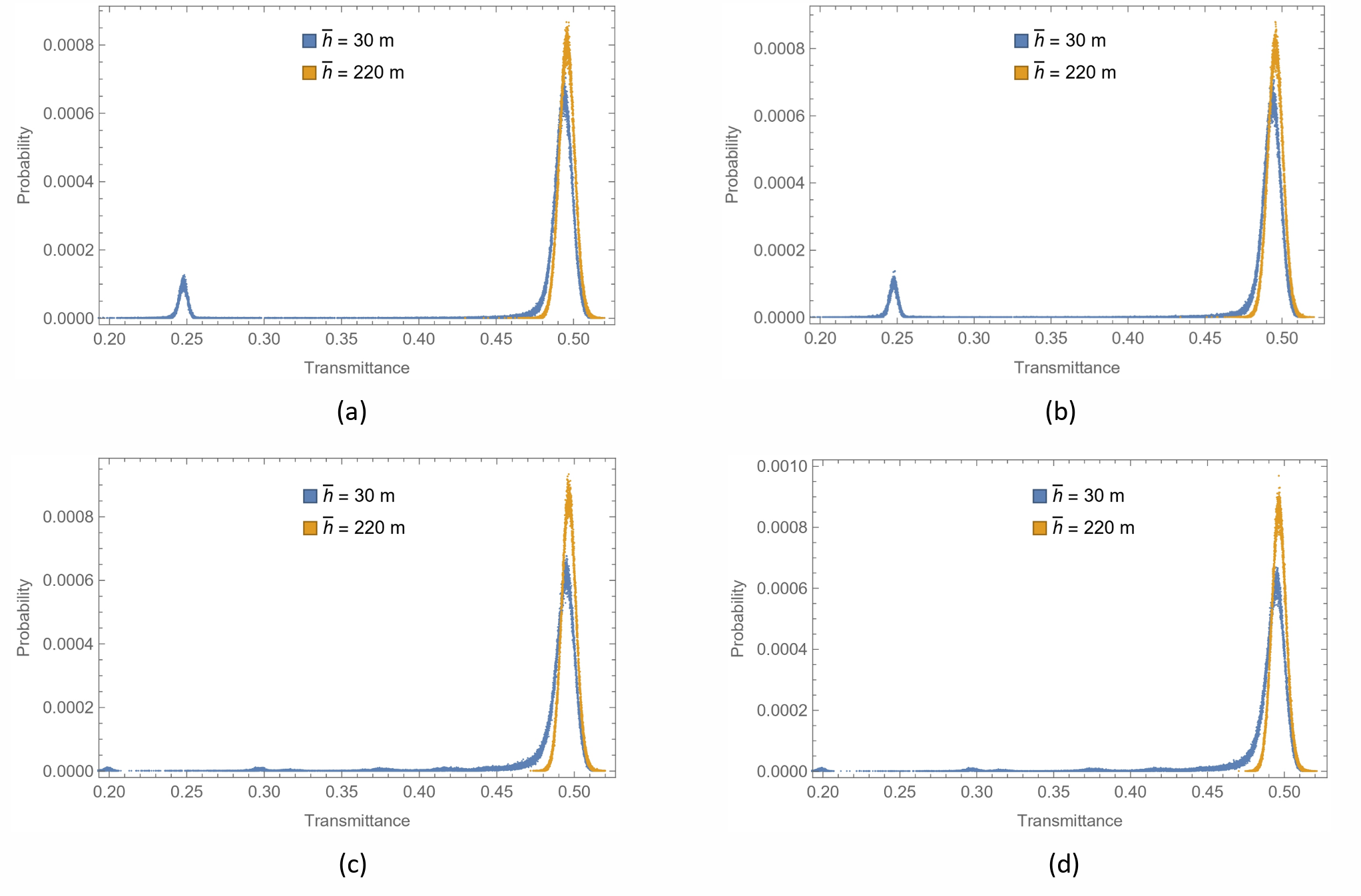}
\caption{(Color online) Plot of PDT variation with different altitude positions at the zenith position for our hybrid model: (a) PDT at day time condition under down-link configuration, (b) PDT at night time condition under down-link configuration, (c) PDT at day time condition under up-link configuration, (d) PDT at night time condition under up-link configuration.}
\label{fig:PDT_DownLink_D_N_UpLink_D_N}
\end{figure*}

\subsubsection{Performance analysis of simulation result}

The proposed hybrid approach primarily relies on short-altitude communication,
employing Gaussian beam-based quantum communication via drones. To
validate the applicability and performance integrity of the proposed
model in the context of FSO communication, we need to analyze the probability
distribution of the transmittance (PDT) of this model. In our analysis,
we appropriately employ both normal and uniform distributions \cite{WHW+18}
for beam parameters ($\boldsymbol{V}$) and incorporate specific optical
values \cite{china-ed} to emulate our model (refer to Table \ref{Parameters-associated-with-link-length}).

To generate PDT plots, we utilize random M5-tuples, generating a substantial
number of values ($10^{6}$ values), and approximate the results to
five decimal places to get well-suited for PDT representation. We
present the transmittance performance in various scenarios encompassing
both up-link and down-link configurations as well as day and night
conditions, at altitudes of $30$ m and $220$ m (refer to Fig. \ref{fig:PDT_DownLink_D_N_UpLink_D_N}). Notably, for the down-link
configuration, the transmittance probability exhibits similar trends
in both day and night conditions (refer to Fig. \ref{fig:PDT_DownLink_D_N_UpLink_D_N}
(a) and \ref{fig:PDT_DownLink_D_N_UpLink_D_N} (b)). At an altitude
of $30$ m, we observe peak transmittance probability values occurring
for transmittance values of about $0.25$ and $0.5$. In this scenario,
the probability distribution is relatively broad when compared to
the $220$ m altitude scenario. Conversely, at $30$ m altitude, the
peak transmittance probability occurs only in the vicinity of a transmittance
value of $0.5$, with a sharply peaked distribution and higher magnitude,
evident in both day and night conditions. In the up-link configuration,
peak transmittance values are consistently located near a transmittance
value of $0.5$ for both day and night conditions (see Fig. \ref{fig:PDT_DownLink_D_N_UpLink_D_N}
(c) and \ref{fig:PDT_DownLink_D_N_UpLink_D_N} (d)). The distribution
nature is broader and slightly lower in value for the $30$ m altitude
compared to the $220$ m altitude scenario. This observation is attributed
to the lower losses incurred at low altitudes ($30$ m), as there
is relatively less interaction with the atmosphere. Conversely, at
high altitudes ($220$ m), the losses are substantial, resulting in
a sharper distribution.

\begin{figure*}[h!]
\centering
\includegraphics[width=\textwidth]{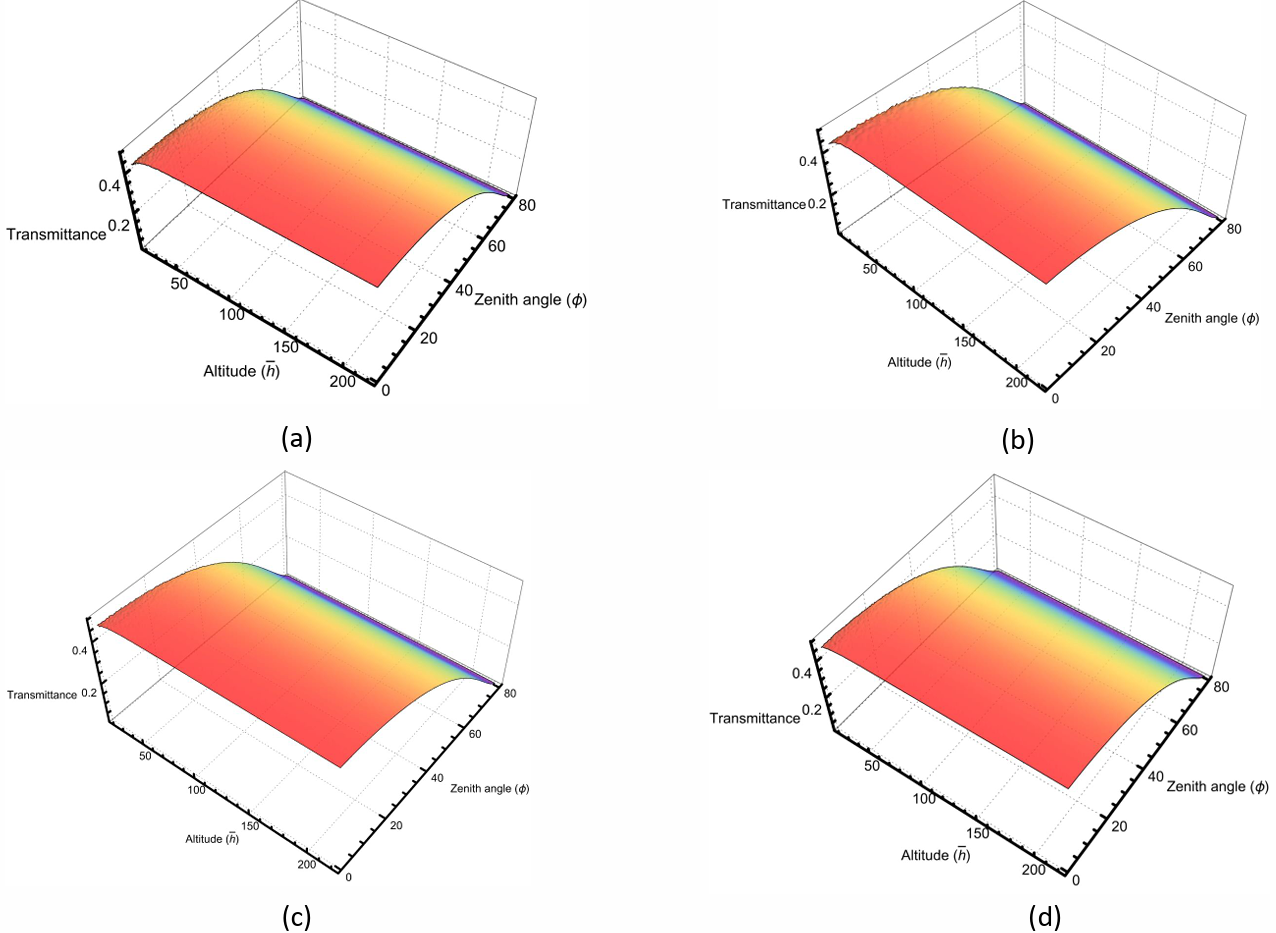}
\caption{(Color online) Variation
of transmittance with altitude ($\overline{h}$) and zenith angle ($\phi$) for our hybrid model: (a) Transmittance at day time condition under down-link configuration, (b) Transmittance at night time condition under down-link configuration, (c) Transmittance at day time condition
under up-link configuration, (d) Transmittance at night time condition under up-link configuration.}
\label{fig:Transmittance_DownLink_D_N_UpLink_D_N}
\end{figure*}

We have also generated plots illustrating the variation in transmittance
concerning altitude ($\overline{h}$) and zenith angle ($\phi$),
as shown in Fig. \ref{fig:Transmittance_DownLink_D_N_UpLink_D_N},
for both up-link and down-link configurations, encompassing both day and night conditions. To generate these plots, we have utilized random sets of M5-tuples, each containing $1000$ values drawn from an appropriate
probability distribution. These random samples of beam parameters allowed us to simulate the transmittance values across various combinations of altitude and zenith angle. Notably, the curvature of the transmittance values across different combinations of altitude and zenith angle exhibits similar trends for all the cases. These findings align with the results obtained from the PDT analysis. It is worth mentioning
that due to the relatively low altitude of the drone-based FSO communication system,
the variation in transmission remains nearly consistent across different
environmental conditions. However, it is important to note that our
hybrid approach can be extended to consider various values of $C_{n}^{2}$
for higher altitudes, as detailed in Appendix A, to gain a deeper
understanding of its applicability under such conditions.

\section{Link configuration, budgeting and margin and time synchronization}
\label{section-4}

\subsection{Link configuration}
\label{link-config}

For longer link distances, it is assumed that the key generation rate of an uplink configuration is roughly one magnitude lower than that of the downlink \cite{Bourgoin:15, maha}, while in the down-link scenario, pointing errors are notably relevant. In the up-link, pointing errors can be mitigated since ground stations can employ more extensive and sophisticated optical systems. However, the turbulence is more concentrated near the earth's surface so for the uplink transmission the turbulence-induced distortion at the beginning significantly increases the beam wandering and divergence angle resulting in a larger channel attenuation as compared to the case of the downlink transmission. 

\begin{figure}[h!]
\begin{minipage}[b]{0.45\textwidth}
\centering
    \includegraphics[width=\textwidth]{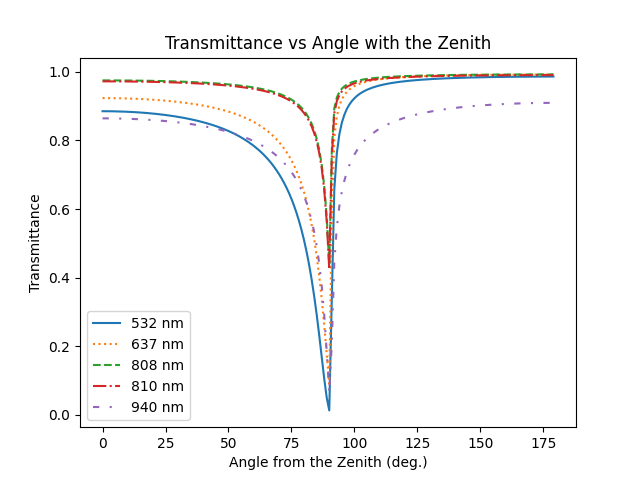}
    \caption{(Color online) Transmittance as a function of different angles with the zenith for different wavelengths.}
    \label{fig:tx-vs-angle}
\end{minipage}
\hspace{0.5cm}
\begin{minipage}[b]{0.45\textwidth}
\centering
    \includegraphics[width=\textwidth, height=5.5cm]{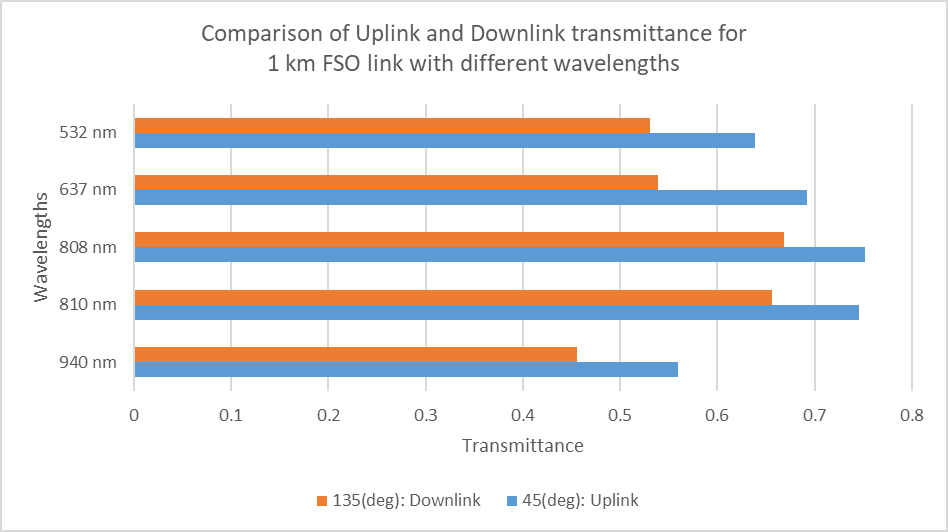}
    \caption{(Color online) Comparison of transmittance for uplink and downlink.}
    \label{fig:up-vs-down}
\end{minipage}
\end{figure}

A comparison of the atmospheric transmittance for a 1 km FSO link as a function of different angles with the zenith for the uplink and downlink configurations with different wavelengths was carried out using MODTRAN software. Fig.\ref{fig:tx-vs-angle} and Fig.\ref{fig:up-vs-down} show the simulated atmospheric transmittance for an urban location with the tropical atmospheric model and 9 km visibility. 

From the simulation results, we can observe that for the shorter links, the transmittance for both uplink and downlink configurations is comparable. And since aerial platforms can fly at much lower altitudes, the total link budget will have minor deviations between the uplink and downlink in terms of geometric loss, atmospheric turbulence, and other types of attenuation.
\subsubsection{Integrated acquisition, pointing, and tracking (APT)}
\begin{figure*}[h!]
\centering
    \includegraphics[width=\textwidth]{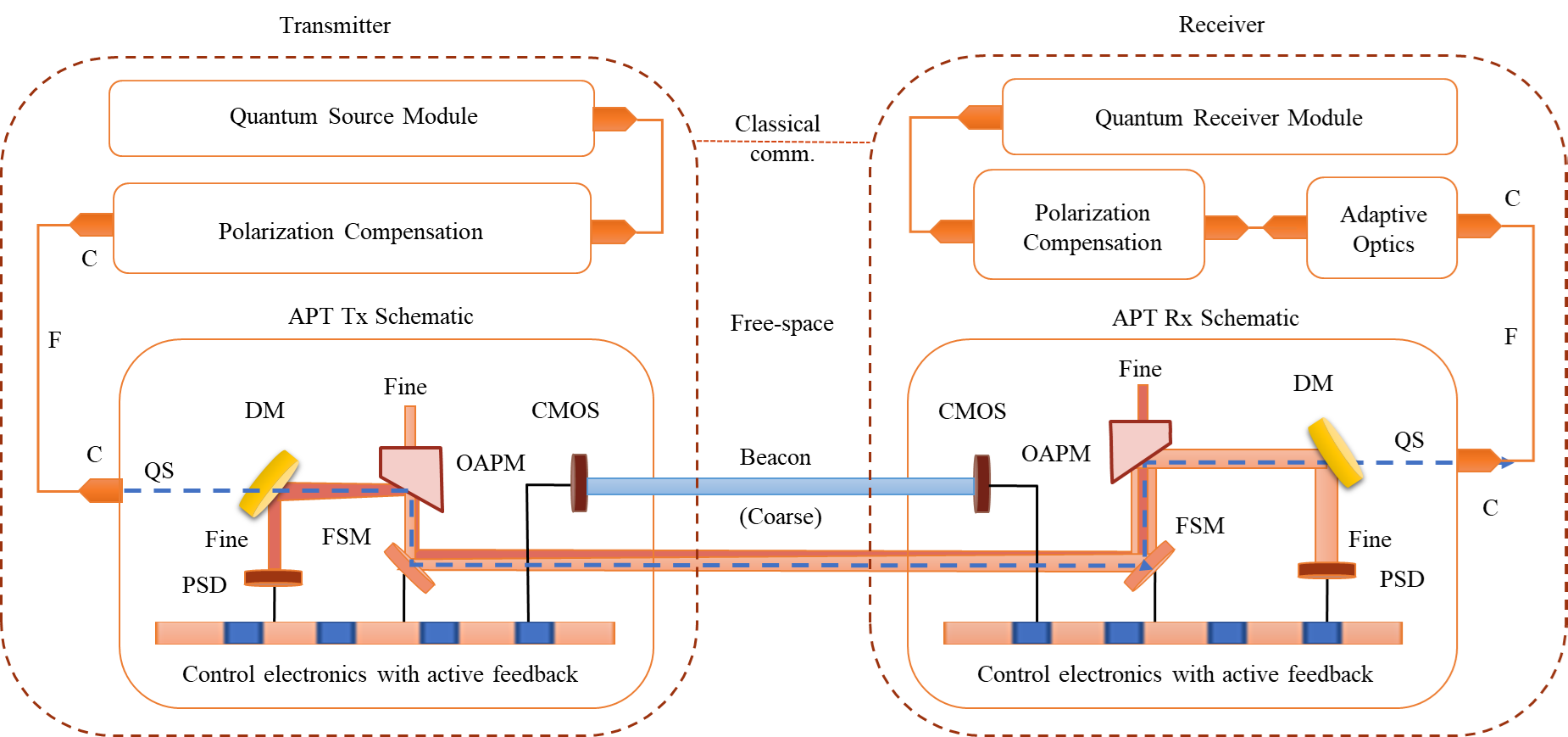}
    \caption{Schematic of integrated acquisition, pointing, and tracking (APT) unit for aerial quantum communication, where the abbreviations used are as follows- C: collimator, F: optical fiber, QS: quantum signal, DM: dichroic mirror, PSD: position sensitive detector, FSM: fast-steering mirror, OAPM: off-axis parabolic mirror, CMOS: camera/sensor.}
    \label{fig:schematic-apt}
\end{figure*}

For aerial quantum communication, distributing the photons simultaneously raises a higher requirement of the dynamically established aerial vehicle-to-ground station links, and to keep the polarization and time series stable during the whole distribution process. Thus there is a need to integrate all the elements for polarization compensation, adaptive optics, collimation, and tracking into an integrated APT unit and perform two-stage tracking, viz. coarse and fine \cite{drone-bb84-2023}. We have presented a high-level architecture of an APT system in Fig. \ref{fig:schematic-apt}.

An APT unit consists of a motorized three-axis (pitch, yaw, and roll) gimbal mount along with a telescope platform. The coarse pointing alignment of the transmitter/receiver telescope is enabled by moving the telescope platform by the gimbal mount using a proportion-integration-differentiation (PID) error signal. This is calculated from the target image using a coaxial zoom camera. The target for this imaging identification is an uncollimated laser beam, typically of the NIR or IR wavelength range on the corresponding receiver or transmitter side.

The telescope on each APT unit collimates light to a beam size optimum for reducing the beam divergence loss, as discussed in Section \ref{Diff-loss}. A carbon-fiber base plate can be used for the telescope platform, where the composite structure design can be optimized for the best thermal stability. Typically 90-degree off-axis parabolic mirror (OAPM) of aperture comparable to the desired beam width is used for collimation. Whereas, the beacon laser beams for the second stage-fine tracking pass through the central hole of the parabolic mirror. The beacon laser has a small aperture, however as it propagates through the link it provides broader FOV, which helps in the coarse tracking. Subsequently, the fine-tracking is performed using a fast-steering mirror (FSM) and a position-sensitive detector (PSD). The PSD is placed at the image position of the dichroic mirror (as shown in Fig. \ref{fig:schematic-apt}). It captures the position of the fine-tracking laser and generates error signals to give feedback to the FSM. Accordingly, FSM aligns itself to reduce this error and achieve tracking with accuracy within the 5 $\mu$m range.

The PSD is mounted at the image position of the transmitter or receiver fiber port to a dichroic mirror (DM). It monitors the focal position of the beacon light to generate the error signal and feedback to the FSM. With proper feedback electronic controls, the transmitter and receiver unit can be pointed at each other within the accuracy of SMF coupling.

APT systems for aerial quantum communication face significant challenges due to the aerial platform-induced jitter, vibrations, and the need for precise synchronization. Mechanical vibrations and jitter from aerial platforms can disrupt optical alignment, requiring real-time feedback-based compensation mechanisms like fast steering mirrors. Effective vibration isolation is also crucial, as environmental factors such as wind and atmospheric turbulence also impact the stability. Moreover, there are constraints with the SWaP (size, weight, and power) factors, as the payload needs to be lightweight and power-efficient for aerial deployment. Overcoming these challenges demands advanced technology and robust testing to maintain a stable optical link while minimizing the system's physical footprint.
\subsection{Link budgeting}
A link budget aims to calculate and analyze the overall performance of a communication link or system, mainly to figure out what distances one could reach with given equipment and to determine whether additional power is available for FSO links under given atmospheric conditions, especially in wireless communication. 

QKD systems rely on optical communications link analysis to have enough photons arriving at the receiver. The main factors that must be considered regarding optical communications are the distance between the transmitter and the receiver, the operating wavelength, all the losses related to atmospheric conditions, geometrical losses, channel turbulence, background noise, and optical losses.

Link budget calculates the minimum power or signal strength required for a communication link to function under specific conditions. In contrast, the link margin represents the additional power or signal strength added to ensure reliability. The link margin is directly related to the link budget.

\subsubsection{Link margin}
The link margin is the gap between the actual received power and the minimum required received signal level.
\begin{equation}
    \text{Link Margin}= P_{\rm t} - A_{\rm tx} - 20\hspace{0.4mm}\text{log} \left( {\frac{\sqrt{2}L\theta_{div}}{\rm D}} \right) - A_{\rm rx} - \alpha_{\rm fog}L - S_{\rm r}
\end{equation}
where, $P_{\rm t}$ is the transmitted power, $A_{\rm tx}$ is the coupling losses at the transmitter, \textit{L} is the range of the FSO link, $\theta_{div}$ is the half-angle divergence, $\alpha_{\rm fog}$ is the attenuation losses due to moisture and $S_{\rm r}$ is the sensitivity of the receiver.

It is imperative that the link margin remains positive, and efforts should be directed toward its maximization. If the link margin becomes negative, the FSO link will no longer be operational.

\begin{figure}[h!]
\centering
    \includegraphics[width=0.5\textwidth]{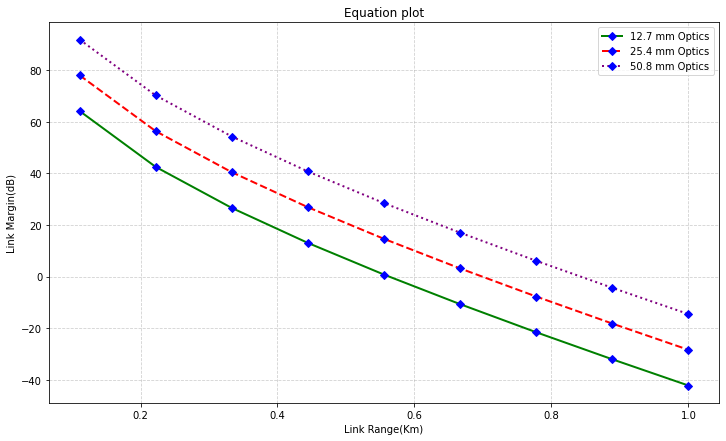}
    \caption{(Color online) Link margin versus link range with various aperture diameters of the receiver lens}
    \label{fig: Link Margin}
\end{figure}

In Fig. \ref{fig: Link Margin}, we have simulated the link margin as a function of link range with various aperture diameters of the receiver lens. It was observed that with increasing distances, the link margin decreases. However, as we increase the aperture diameter of the receiving optics, the link margin increases.
\subsection{Time synchronization}
Time synchronization is essential to provide a time reference that allows two distant users to generate the correlated information simultaneously. Generally, components like lasers, modulators, and detectors can introduce jitter due to their finite response times and inherent noise. Precise compensation for this jitter can be mitigated by the use of stable and precise reference clocks, the implementation of delay compensation techniques, and the use of high-quality optical components with low jitter, which is necessary to ensure accurate synchronization. For aerial quantum communication, the distance between the transmitter and receiver continuously changes; hence, time synchronization is implemented in a particular manner.

A fault-tolerant synchronization based on de Bruijn sequences is suitable for timing and synchronization over high-loss space-to-ground communication channels. It provides an efficient sequence position encoding, which exploits achieving robustness to beacon corruption in the decoding process \cite{Hybrid}.

A fiber optic two-way quantum clock synchronization combined with microwave frequency transfer technology gives picosecond scale synchronization precision, which promises femtosecond precision over intercity optical fiber links in the future \cite{Two-way}.

Qubit-based synchronization (Qubit4sync) with a cross-correlation scheme is a synchronization procedure that only needs the same photons encoding the quantum state exchanged in  QKD protocol. This avoids additional hardware, makes it cheaper, and lowers failure probability due to hardware \cite{Qubit4sync}.

Qubit-based clock synchronization using the Bayesian probabilistic algorithm efficiently finds the clock offset without sacrificing the secure key. In comparison with other protocols, it is more robust to channel loss, noise, and clock drift \cite{Bayesian}.

In satellite-to-ground large-distance, quantum communication where independent reference clocks are employed GPS pulse-per-second (PPS) signal and an assistant pulse laser are used for time synchronization \cite{Yin_2012}.

In 2021, the Space Application Centre (SAC) of ISRO used a novel synchronization technique enabled with NavIC for a distance of 300 m to achieve a secure key rate of 300 kbps \cite{Time-sync}.
\section{Simulation of quantum teleportation using entanglement swapping through a swarm of drone network} \label{aqns}
In this section, we have presented a use case where we simulate quantum teleportation between two distant nodes using entanglement swapping through a swarm of drones. We have performed the simulation using the Network Simulator for Quantum Information using Discrete events (NetSquid). NetSquid \cite{netsquid} is a software tool for the modeling and simulation of scalable quantum networks developed by QuTech. This QN simulation directs towards a software-defined networking (SDN)-based architecture \cite{netsquid-q-tel} to manage the distribution of end-to-end entangled pairs between two ground stations (GSs). The architecture is adaptable for quantum computing and QKD services. 

In the simulation scheme presented in  Fig. \ref{fig:netsquid-sdn}, a swarm of drones comprising of $n$ quantum repeaters (QR), designated as $D^{QR}_n$, is distributed between two end stations performing the quantum teleportation. The drones nearest to the end stations, Alice and Bob can be referred to as $D^{QR}_1$ and $D^{QR}_n$. We consider that each QR drone has quantum memory (QM), which can house two quantum particles entangled with the adjacent neighboring QR drones' particles. When the QR drone performs a Bell state measurement (BSM) on its two quantum particles, the measurement will result in the entanglement swapping amongst the two neighboring QR drone particles. The entire scheme is discussed in detail below:

\begin{figure}[h!]
\begin{minipage}[b]{0.6\textwidth}
\centering
    \includegraphics[width=\textwidth, height= 5cm]{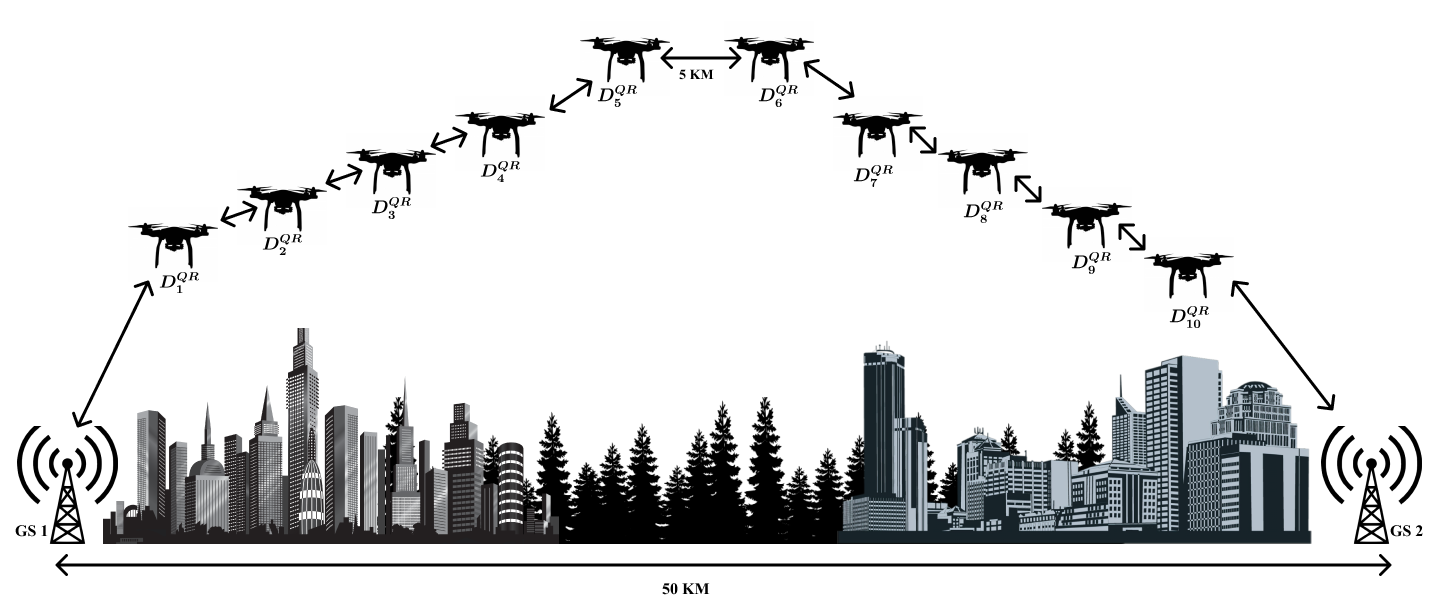}
    \caption{Scheme for the quantum teleportation using entanglement swapping using a swarm of drones \cite{netsquid-q-tel}.}
    \label{fig:netsquid-sdn}
\end{minipage}
\hspace{0.3cm}
\begin{minipage}[b]{0.4\textwidth}
\centering
    \includegraphics[width=\textwidth]{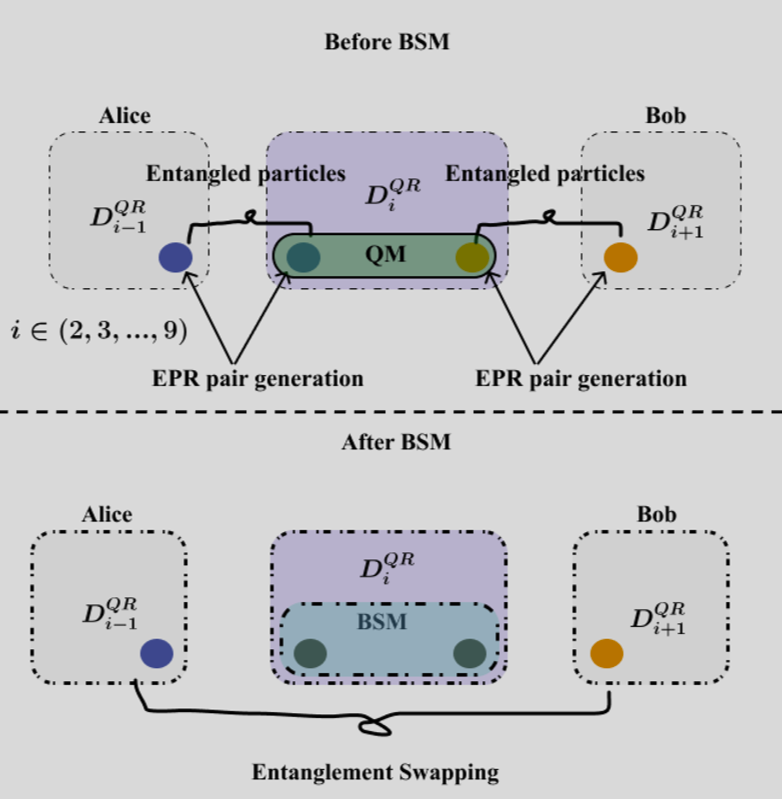}
    \caption{Entanglement swapping \cite{netsquid-ent-swap}.}
    \label{fig:netsquid-ent-swap}
\end{minipage}
\end{figure}

\begin{enumerate}
    \item [i] A swarm of $n$ QR drones ($D^{QR}_1$ to $D^{QR}_n$), is distributed between the two end stations performing quantum teleportation, Alice and Bob. 
    \item [ii] Each QR drone ($D^{QR}_i$) possesses two particles, each entangled with the two subsequent neighboring QR drones' particles ($D^{QR}_{i-1}$ and $D^{QR}_{i+1}$). The entangled pairs may be stored on the QR drones using quantum memories before the take-off or distributed in real-time (refer to Fig. \ref{fig:netsquid-ent-swap}). 
    \item [iii] The end stations, say Alice and Bob share an entangled pair with the $D^{QR}_1$ and $D^{QR}_n$, respectively.
    \item [iv] Quantum entanglement swapping is executed at $D^{QR}_1$ resulting in the entanglement between Alice and $D^{QR}_2.$
    \item [v] In this way, the entanglement swapping \cite{netsquid-ent-swap} is repeated consequently for the rest of the QR drone chain, from the $D^{QR}_2$ to $D^{QR}_n$. At the end after $n$ entanglement swapping, Alice's particle gets entangled with Bob's particle. 
    \item [vi] After the establishment of an entanglement pair between Alice and Bob, for the quantum teleportation Alice performs a complete measurement of the \textit{von Neumann} type on the joint system, consisting of her particle from the shared EPR pair and the \textit{arbitrary unknown state} ($\ket{\psi}$) particle whose information needs to be shared.
    \item [vii] She then sends the outcome of her measurement to Bob through the classical channel, who then applies the required unitary (rotation) operations on his EPR particle to receive  $\ket{\psi}$. Hence, the state is teleported from Alice's lab to Bob's lab.
\end{enumerate}

The simulation of the above quantum teleportation scheme was carried out for different configurations on NetSquid. We have calculated the fidelities of the resulting teleported states and performed the time analysis for the execution of the entire scheme. In the Node-to-Node configuration, quantum teleportation between Alice and Bob separated at a $5$ km distance without any intermediate QR drone was carried out. While in the End-to-End configuration, quantum teleportation over $50$ km distance using entanglement swapping as per the above scheme, through ten QR drones, each separated at $5$ km distance between Alice and Bob was carried out. The results are shown in the Table \ref{Table:NetSquid}.

\begin{table}[h]
\centering
\begin{tabular}{c c c} 
\hline
Parameters & Node-to-Node & End-to-End \\ 
\hline
Fidelity & 0.964 &  0.1516   \\ 
Time (ns)  & 5 &  236111 \\ 
\hline
\end{tabular} \vspace{3pt}
\caption{Simulation of quantum teleportation using entanglement swapping on NetSquid for different configurations.}
\label{Table:NetSquid}
\end{table}
\section{Conclusion}
\label{conc}
In this work, we have emphasized the necessity and importance of non-terrestrial platforms for future quantum communication which will explore free-space mediums in an optimal way to provide end-to-end solutions. We have attempted to adequately address the challenges of aerial quantum communication. We have introduced a hybrid model that elaborates on the characteristics of transmittance with the variation of zenith angle in densely humid medium and low altitude signal transmission. Further, we have analyzed the average visibility of Pune city for the last two years for a feasibility study to implement aerial quantum communication using Drones. Finally, we have simulated quantum teleportation between two distant nodes via a swarm of quantum drone networks utilizing QSDN. The SDN technology will have a significant role in near-future integrated quantum networks and services. Our work aims to stimulate further research and explore the boundaries in this promising field.

\section*{Acknowledgements}
The authors acknowledge the support from R\&D IT, MeitY, India. 

We also thank Ms. Akshara Jayanand Kaginalkar, C-DAC, Pune for the availability of the meteorological data.
\bibliographystyle{unsrt}
\bibliography{reference}

\section*{Appendix A \label{sec:Appendix-A}}

The refractive index structure constant $C_{n}^{2}$ is entirely reliant
on the specific atmospheric conditions, which can be highly intricate
and multifaceted. To be more precise, it is contingent upon the altitude
within the Earth's atmosphere where the space communication system
operates, such as a low Earth orbit (LEO) satellite or a drone. Numerous
models have been proposed in the past (see Ref. \cite{LC06})
to characterize this dependence. In our analysis, focusing on quantum
communication at lower altitudes, we have opted for the SLC-D model \cite{GBM+88}, which is parameter-free and has been extensively employed in densely humid weather conditions, thick cloud cover, and underwater environments. It is worth noting that such models are customarily tailored for specific geographic sites and subtropical climates. The SLC-D model takes the following mathematical form \cite{LC06},

\[
\begin{array}{lcl}
C_{n}^{2}\left(\overline{h}\right) & = & \begin{cases}
\begin{array}{c}
1.70\times10^{-14}\\
\frac{3.13\times10^{-13}}{\overline{h}}\\
1.30\times10^{-15}\\
\frac{8.87\times10^{-7}}{\overline{h}^{3}}\\
\frac{2.00\times10^{-16}}{\overline{h}^{0.5}}
\end{array} & \begin{array}{c}
\overline{h}<18.5\,{\rm m}\\
18.5\,{\rm m}<\overline{h}<240\,{\rm m}\\
240\,{\rm m}<\overline{h}<880\,{\rm m}\\
880\,{\rm m}<\overline{h}<7,200\,{\rm m}\\
7,200\,{\rm m}<\overline{h}<20,000\,{\rm m}
\end{array}.\end{cases}\end{array}
\]
A comparable model is accessible for night time condition. An alternative iteration of the SLC-D profile model can be found in published articles \cite{PS94,RW18}, outlined as follows:

\[
\begin{array}{lcl}
C_{n}^{2}\left(\overline{h}\right) & = & \begin{cases}
\begin{array}{c}
0\\
\frac{4.008\times10^{-13}}{\overline{h}^{1.054}}\\
1.30\times10^{-15}\\
\frac{6.352\times10^{-7}}{\overline{h}^{2.966}}\\
\frac{6.209\times10^{-16}}{\overline{h}^{0.6229}}
\end{array} & \begin{array}{c}
0\,{\rm m}<\overline{h}<19\,{\rm m}\\
19\,{\rm m}<\overline{h}<230\,{\rm m}\\
230\,{\rm m}<\overline{h}<850\,{\rm m}\\
850\,{\rm m}<\overline{h}<7,000\,{\rm m}\\
7,000\,{\rm m}<\overline{h}<20,000\,{\rm m}
\end{array}.\end{cases}\end{array}
\]

\end{document}